# Drift-diffusion model for single layer transition metal dichalcogenide field-effect transistors

David Jiménez

Departament d'Enginyeria Electrònica, Escola d'Enginyeria,
Universitat Autònoma de Barcelona, 08193-Bellaterra, Spain

david.jimenez@uab.es

*Dated 12-July-2012*

**Abstract**

A physics-based model for the surface potential and drain current for monolayer transition metal dichalcogenide (TMD) field-effect transistor (FET) is presented. Taking into account the 2D density-of-states of the atomic layer thick TMD and its impact on the quantum capacitance, a model for the surface potential is presented. Next, considering a drift-diffusion mechanism for the carrier transport along the monolayer TMD, an explicit expression for the drain current has been derived. The model has been benchmarked with a measured prototype transistor. Based on the proposed model, the device design window targeting low-power applications is discussed.

**Introduction**

A great deal of interest in two-dimensional materials analogues of graphene has appeared among the scientific community since the demonstration of isolated 2D atomic plane crystals from bulk crystals [1]. Dimensionality is key for the definition of material properties and the same chemical compound can exhibit dramatically different properties depending on whether it is arranged in dots (0D), wires (1D), sheets (2D) or bulk (3D) crystal structure. Notably, experimental studies of 2D atomic crystals were lacking until recently because of the difficulty in their identification [1]. Representative of this class are the 2D monolayer of transition metal dichalcogenides (TMD) with a chemical formula $MX_2$, where M stands for a transition metal and X for Se, S, or Te. The potential of this family of layered materials for flexible electronics was proposed by Podzorov et al., who demonstrate an ambipolar $WSe_2$ p-FET with high hole mobility (~500 $cm^2$/V-s) [2]. The electronic properties of TMDs vary from semiconducting (e.g., $WSe_2$) to superconducting (e.g., $NbSe_2$). The semiconducting monolayer TMDs, like $MoS_2$, $MoSe_2$, $MoTe_2$, $WS_2$, and $WSe_2$ are predicted to exhibit a direct gap in the range of 1–2 eV [3]. The wide gap together with a promising ability to scale to short gate lengths because of the optimum electrostatic control of the channel, by virtue of its thinness, make monolayer TMDs very promising for low power switching and optoelectronics applications. The first 2D crystal based FET was demonstrated using a monolayer $MoS_2$ as the active channel [4]. Low power switching with an $I_{ON}/I_{OFF}$~$10^8$ and subthreshold swing (SS) of 74 mV/decade at room temperature, was experimentally measured. More recently, a monolayer p-type $WSe_2$ FET with an optimum SS ~ 60 mV/decade and $I_{ON}/I_{OFF}$ >$10^6$ was demonstrated [5].

To boost the development of 2D-material based transistor technology, modeling of the electrical characteristics is essential to cover aspects as device design optimization, projection of performances, and exploration of low-power switching circuits. Some models aimed to explore the performance limits of monolayer TMD transistors have been reported assuming ballistic transport [6,7]. However, the behavior of state-of-the art devices is far from ballistic and a diffusive transport regime seems more appropriate. In this context, I propose a model for the current-voltage (I-V) characteristics of monolayer TMD FETs, based on the drift-diffusion theory. As a previous step a surface potential model, accounting for the 2D density-of-states (DOS$_{2D}$) of monolayer TMDs, is proposed. The DOS$_{2D}$ has a profound impact on the quantum capacitance, which is essentially different from that of a nanowire (1D) or a bulk (3D) material. Analytical expressions have been derived for both the surface potential and drain current covering both subthreshold and above threshold operation regions.

**Surface potential model**

Let us consider a dual-gate monolayer TMD FET with the cross-section depicted in the inset of Fig. 1a. It consists of one atomic layer thick TMD playing the role of the active channel. The source and drain electrodes contact the monolayer TMD and are assumed to be ohmic. The electrostatic modulation of the carrier concentration in the 2D sheet is achieved via a double-gate stack consisting of top and bottom gate dielectric and the corresponding metal gate. The source is grounded and considered the reference potential in the device. The electrostatics of this device can be understood using the equivalent capacitive circuit depicted in Fig. 2. Here, $C_t$ and $C_b$ are the top and bottom oxide capacitances and $C_q$ represents the quantum capacitance of the 2D sheet. The charge density (per unit area) is calculated by integrating the DOS$_{2D}$ over all the energies and can be expressed as:

$$Q_c = Q_p + Q_n = q \int_{-\infty}^{0} DOS_{2D}(E) f(E_F - E) dE - q \int_{0}^{\infty} DOS_{2D}(E) f(E - E_F) dE \quad (1)$$

where $Q_p$ and $Q_n$ refer to the positive (holes) and negative (electrons) charge contributions, respectively; f(E) is the Fermi-Dirac function, and $E_F = qV_c$ is the Fermi level. The parameter $V_c$ represents the voltage drop across $C_q$ or surface potential. For the sake of getting a simple model f(E)~1 for E<$E_F$ and f(E)~exp(($E_F$-E)/kT) has been assumed. Noting that $DOS_{2D}(E) = D_0 \sum_n H(E - E_n)$, with $D_0 = \frac{m^*}{\pi \hbar^2}$, where m* is the effective mass, $E_n$ represents the energy of the n$^{th}$-subband, H(E) is the Heaviside function, and considering that the ground state (n=0) is the more relevant in determining the carrier density, then Eq. (1) can be written as:

$$Q_p = -q^2 D_0 V_c - q D_0 (E_0 - kT); \quad Q_n = -q D_0 kT e^{\frac{qV_c - E_0}{kT}}, \quad qV_c \leq -E_0$$

$$Q_p = q D_0 kT e^{\frac{-qV_c - E_0}{kT}}; \quad Q_n = -q D_0 kT e^{\frac{qV_c - E_0}{kT}}, \quad q|V_c| < E_0$$

$$Q_p = q D_0 kT e^{\frac{-qV_c - E_0}{kT}}; \quad Q_n = -q^2 D_0 V_c + q D_0 (E_0 - kT), \quad qV_c \geq E_0 \quad (2)$$

where $E_0=E_g/2$ and $E_g$ is the gap of the monolayer TMD. From Eq. (2), the quantum capacitance defined as $C_q$=-$dQ_c/dV_c$, results in:

$$C_q = C_{q,p} + C_{q,n} = q^2 D_0 + q^2 D_0 e^{\frac{qV_c - E_0}{kT}}, \quad qV_c \leq -E_0$$

$$C_q = C_{q,p} + C_{q,n} = q^2 D_0 e^{\frac{-qV_c - E_0}{kT}} + q^2 D_0 e^{\frac{qV_c - E_0}{kT}}, \quad q|V_c| < E_0$$

$$C_q = C_{q,p} + C_{q,n} = q^2 D_0 e^{\frac{-qV_c - E_0}{kT}} + q^2 D_0, \quad qV_c \geq E_0 \quad (3)$$

Under nonequilibrium conditions ($V_{ds} \neq 0$), a single Fermi level cannot be assumed. Instead, two distinct quasi-Fermi levels for computing the electron $V_n(x)$ and hole $V_p(x)$ concentrations and currents have to considered. Here *x* denotes the transport direction. In this work I consider the modeling of unipolar p-FETs. Extension to unipolar n-FETs is straightforward. Coming back to Fig. 2, $V_p(x)$ is zero at the source end (x=0) and $V_{ds}$ at the drain end (x=L). Applying basic circuit laws to the equivalent capacitive network, the following relation can be obtained:

$$V_c(x) = \frac{Q_p(V_c)}{C_t+C_b} + \left(V_{gs} - V_{gs0} - V_p(x)\right)\frac{C_t}{C_t+C_b} + \left(V_{bs} - V_{bs0} - V_p(x)\right)\frac{C_b}{C_t+C_b} \quad (4)$$

where $V_{gs}$-$V_{gs0}$ and $V_{bs}$-$V_{bs0}$ are the top and back gate-source voltage overdrive, respectively. These quantities comprise work-function differences between the gates and the TMD monolayer, eventual charged interface states at the TMD monolayer/oxide interfaces, and intentional or unintentional doping of the TMD monolayer.

**Drain current model**

To model the drain current of a monolayer TMD p-FET a drift-diffusion transport is assumed under the form $I_{ds}$=-$WQ_p(x)v(x)$, where W is the gate width, and v(x) the hole drift velocity v=$\mu$E, where E is the electric field and $\mu$ is the hole effective mobility. Applying E=-d$V_p(x)$/dx, inserting the above expression for v, and integrating the resulting equation over the device length, the drain current becomes $I_{ds} = \mu \frac{W}{L} \int_0^{V_{ds}} Q_p dV_p$. In order to get an explicit expression for the drain current, the integral is solved using $V_c$ as the integration variable consistently expressing $Q_p$ as a function of $V_c$ :

$$I_{ds} = \mu \frac{W}{L} \int_{V_{cs}}^{V_{cd}} Q_p \frac{dV_p}{dV_c} dV_c \quad (5)$$

where $V_{cs}$ is obtained from Eq. (4) as $V_c(V_p$=0$)$. Similarly, $V_{cd}$ is determined as $V_c(V_p$=$V_{ds})$. Moreover, Eq. (4) provides the relation $\frac{dV_p}{dV_c} = -\left(1 + \frac{C_{q,p}}{C_t+C_b}\right)$, where $C_{q,p}$=-$dQ_p$/$dV_c$. Inserting

this expression into Eq. (5), the following explicit drain current expression can be finally obtained:

$$I_{ds} = \mu \frac{W}{L} \{g(V_c)\}_{V_{cs}}^{V_{cd}}$$

$$g(V_c) = \left(1 + \frac{q^2 D_0}{C_t + C_b}\right)\left(\frac{q^2 D_0}{2} V_c^2 + q D_0 (E_0 - kT) V_c\right), \quad qV_c < -E_0$$

$$g(V_c) = D_0(kT)^2 \left(e^{\frac{-qV_c - E_0}{kT}} + \frac{q^2 D_0}{C_t + C_b} \frac{1}{2} e^{\frac{-2(qV_c + E_0)}{kT}}\right), \quad qV_c \geq -E_0 \quad (6)$$

where g($V_c$) takes different forms whether $qV_c < -E_0$ (above threshold region) or $qV_c \geq -E_0$ (subthreshold region). To take into account eventual saturation velocity effects, the physical channel length should be replaced by an effective length $L_{eff} = L + \mu \frac{|V_{ds}|}{v_{sat}}$, where $v_{sat}$ is the hole saturation velocity.

To test the model I have benchmarked the resulting I-V characteristics with experimental results reported in Ref. [5], which are unipolar p-type FETs with a channel consisting of a monolayer $WSe_2$. The channel was contacted with Pd contacts acting as a source and drain electrodes. The p-type conduction is likely to the small barrier height to the valence band at the Pd-$WSe_2$ interface. The gap of the monolayer $WSe_2$ is $E_g \sim 1.68$ eV and the valence band effective mass along the transport direction is calculated to be m*=0.64$m_0$ (K → Γ), $m_0$ being the free electron mass, estimated from the dispertion relations [3]. The device under test has L=9.4 μm, W=1 μm, top dielectric is $ZrO_2$ of 17.5 nm and relative permittivity ∼ 12.5, and the bottom dielectric is silicon oxide of 270 nm. The backgate voltage was -40 V. The flat-band voltages $V_{gs0}$ and $V_{bs0}$ were tuned to -0.5 V and 0 V, respectively, to provide an appropriate shift of the transfer characteristics according to the experiment. A constant hole effective mobility of 250 cm²/V-s was assumed, consistent with measurements. A source/drain resistance of 300 Ω provides a good fit with the experiment. The resulting I-V transfer and output characteristics are shown in Fig. 1. In accordance with the experiment a SS ~ 60 mV/decade at room temperature and $I_{ON}/I_{Off}$ ~$10^6$ is predicted by the model (Fig. 1a). Note that no interface trap capacitance ($C_{it}$) was needed to be included in the model to match the experiment because the near ideal subthreshold slope suggests that $C_{it} \ll C_{ox}$. The output characteristics show saturation-like behavior at high $V_{ds}$ (Fig. 1b). Saturation velocity effects are not expected relevant for this transistor because $\frac{\mu}{v_{sat}} \sim 2.5\ nm/V$, giving $L_{eff} \approx L$. At low $V_{ds}$ the model nicely reproduces the observed linear behavior, indicative of ohmic metal contacts. The agreement between the proposed model (solid lines) and the experiment (symbols) demonstrates its accuracy. Next, by using the model, the tradeoff between $I_{ON}$ and $I_{ON}/I_{Off}$ is calculated (Fig. 3). Ten orders of magnitude between switching states could be achieved, although at the expense of the $I_{ON}$. For the reported transistor, an $I_{ON}/I_{Off}$ ~$10^6$ with $I_{ON}$ ~1 μA/μm @ power supply voltage $V_{DD}$=0.6 V could be achieved. Nevertheless, a huge improvement of the $I_{ON}$ may be possible via channel length scaling. A simulation of an hypothetical transistor of L=100 nm assuming the same hole mobility as the reference transistor gives a factor x100 of ON-current improvement for a fixed $I_{ON}/I_{Off}$.

In conclusion, a surface potential and drain current model for monolayer TMD transistors has been proposed, taking into account the 2D semiconducting nature of monolayer TMDs. The drain current is formulated assuming a drift-diffusion theory, which seems appropriate for explaining the experimental results of reported devices till date. These transistors hold promise for low-power switching applications. The proposed model should be valid for other transistors relying on 2D atomic layer thick channels.

ACKNOWLEDGMENTS

I acknowledge the funding of the Ministerio de Ciencia e Innovación under contract FR2009-0020 and TEC2009-09350, and the DURSI of the Generalitat de Catalunya under contract 2009SGR783

FIGURE CAPTIONS

Fig. 1. Transfer (a) and output (b) characteristics obtained from the analytical model (solid lines) compared with experimental results from Ref. [5] (symbols). Inset: cross section of the dual-gate monolayer TMD transistor.

Fig. 2. Equivalent capacitive circuit of the dual-gate monolayer TMD transistor.

Fig. 3. Tradeoff between ON-current and ON-OFF current ratio

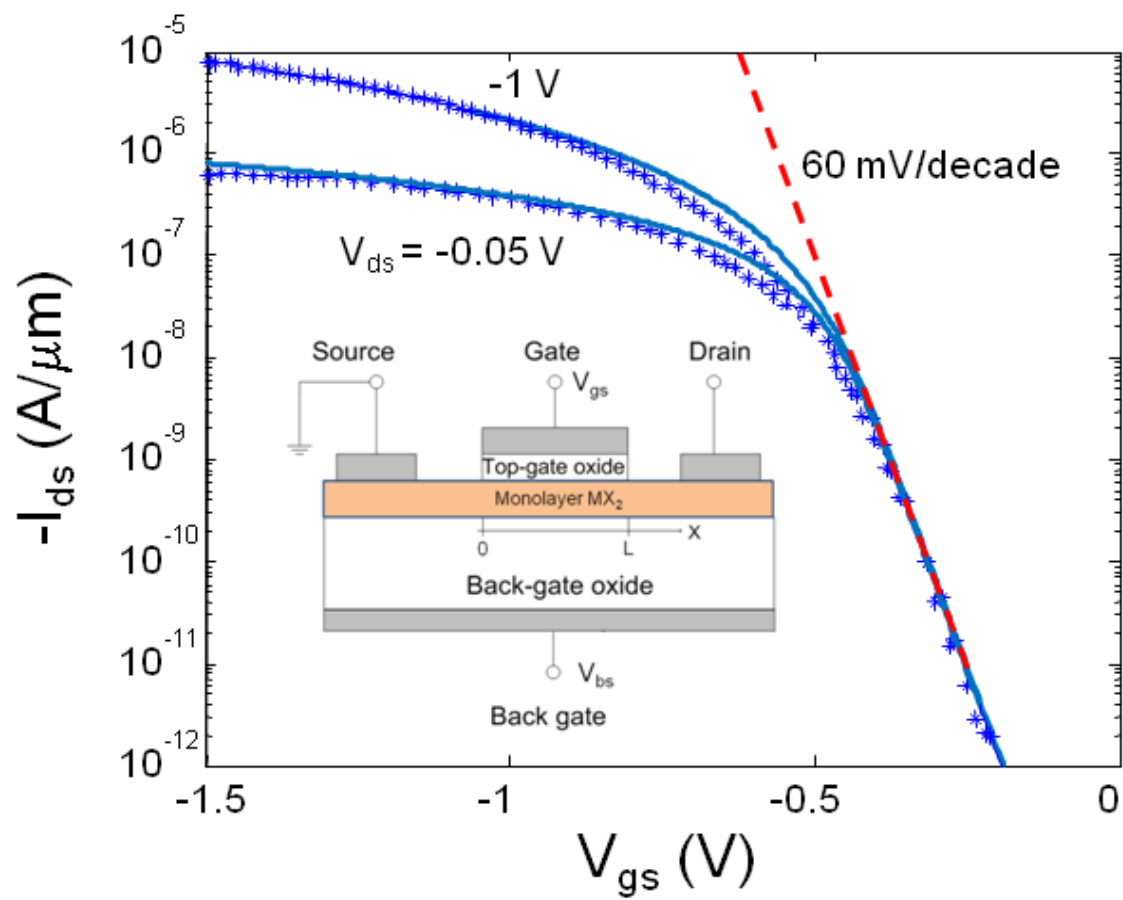

Figure 1a

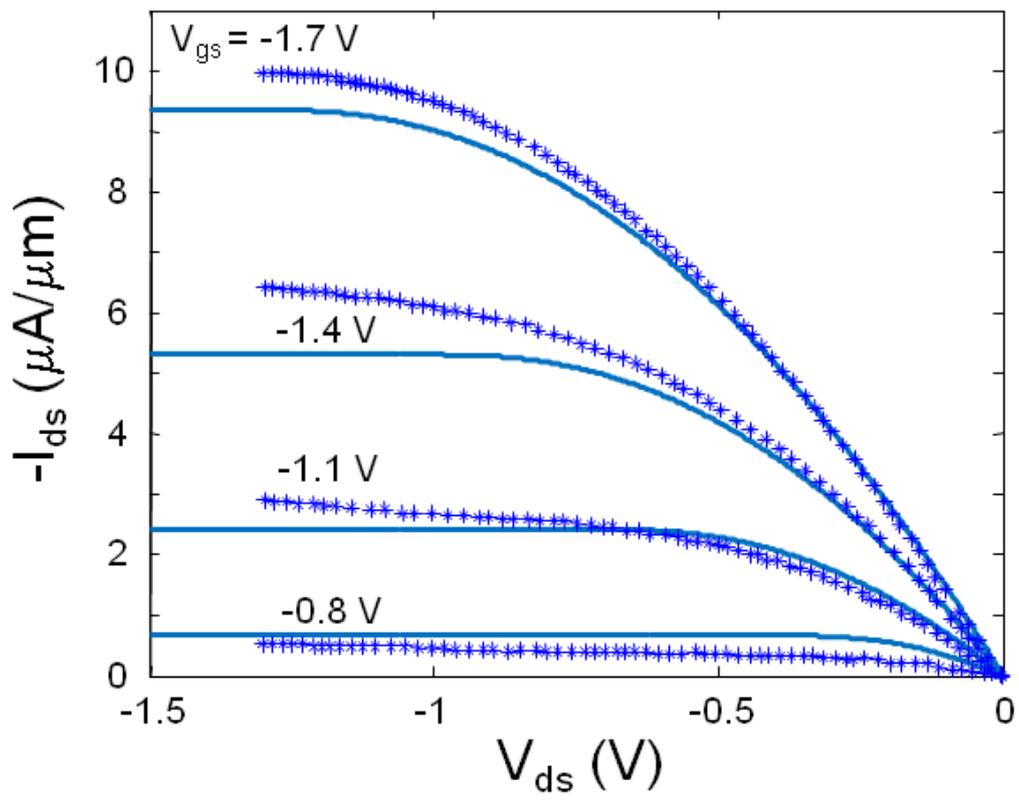

Figure 1b

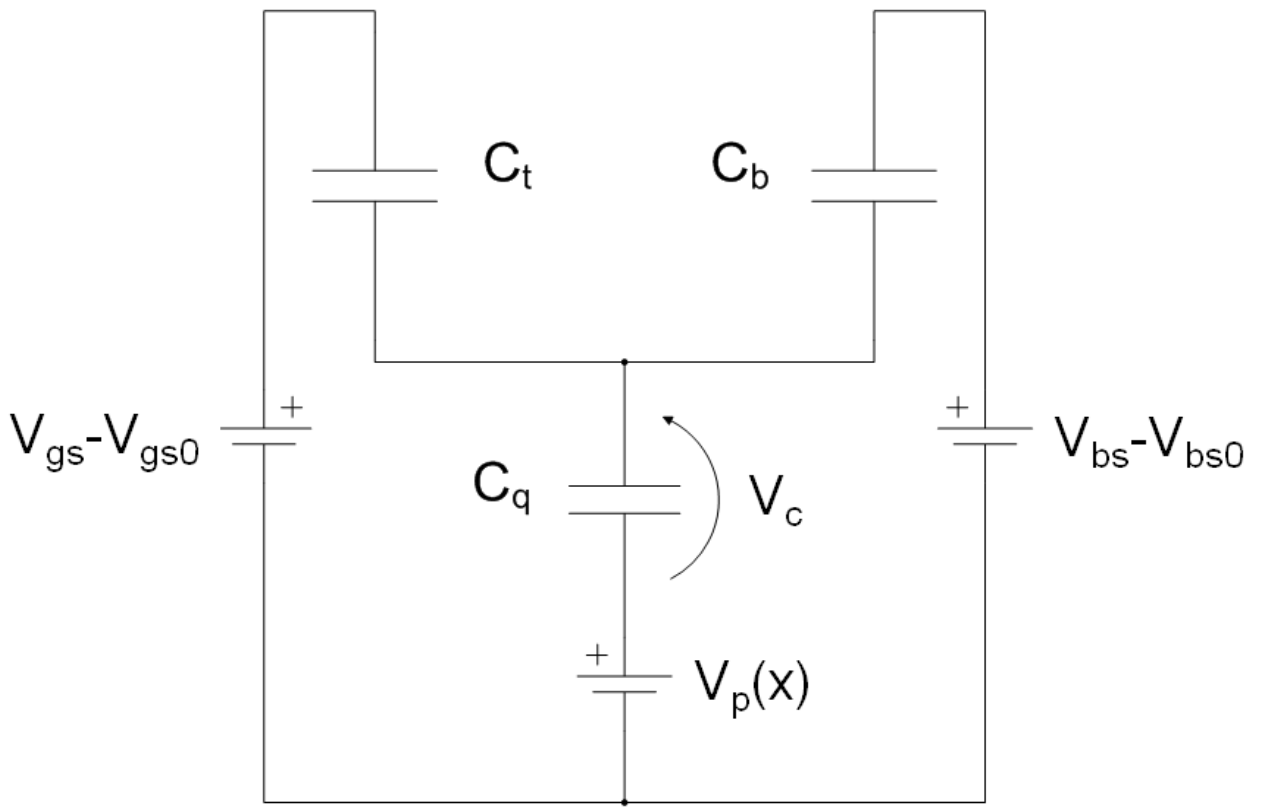

Figure 2

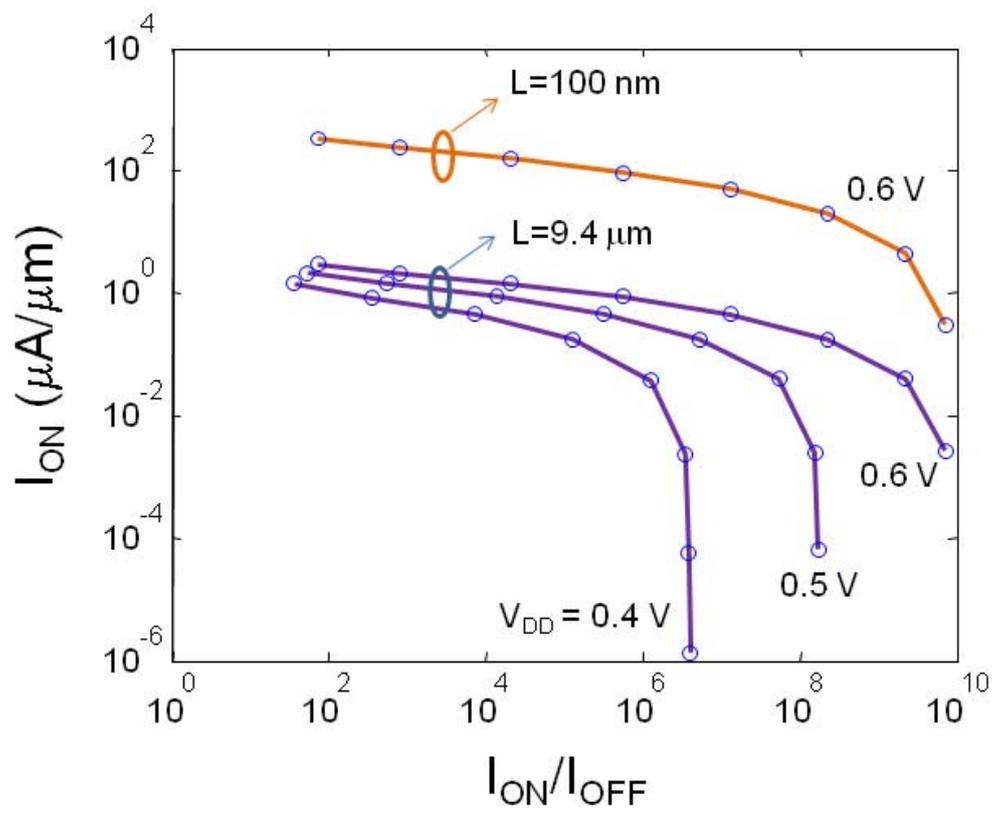

Figure 3